\documentclass[useAMS,usenatbib,usegraphicx]{mn2e}
\usepackage{epsfig}
\usepackage{amsmath}
\usepackage{amssymb}
\def\d{{\mathrm d}}
\def\exp{{\mathrm exp}}

\def\S{{\mathrm S}}

\newcommand{\AAA}[3]    {\mbox{#3, A\&A,~#1,~#2}}

\newcommand{\ApJ}[3]    {\mbox{#3, ApJ,~#1,~#2}}

\newcommand{\AJ}[3]     {\mbox{#3, Astron.~J.,~#1,~#2}}
\newcommand{\MNRAS}[3]  {\mbox{#3, MNRAS,~#1,~#2}}
\newcommand{\Nature}[3] {\mbox{#3, Nature,~#1,~#2}}

\newcommand{\NewARev}[3]{\mbox{#3  New Astron. Rev,~#1,~#2~}}

\newcommand{\PASJ}[3]   {\mbox{#3  PASJ,~#1,~#2}}

\newcommand{\astroph}[1]{\mbox{preprint~(astro-ph/#1)}}

\title{A simple model for the evolution of super-massive black holes
  and the quasar population}

\author[A. Mahmood, J. Devriendt and J. Silk]
  {Asim Mahmood$^1$\thanks{E-mail: mahmood@astro.ox.ac.uk},
   Julien E. G. Devriendt$^{1,2}$
  and Joseph Silk$^1$
 \\
  $^1$Department of Astrophysics, University of Oxford, Keble Road, Oxford, OX1
  3RH, U.K\\
    $^2$Observatoire Astronomique de Lyon, 9 Avenue Charles Andr\'e,
  69561 Saint-Genis Laval cedex, France}

\date{Received ...; accepted ...}

\begin{document}

\label{firstpage}

\maketitle


\begin{abstract}
An empirically motivated model is presented for accretion-dominated growth 
of the super massive black holes (SMBH) in galaxies, 
and the implications are studied for the 
evolution of the quasar population in the universe.  We investigate the 
core aspects of the quasar population, including space density evolution,
evolution of the characteristic luminosity, plausible minimum masses of 
quasars, the mass function of SMBH and their formation epoch 
distribution. Our model suggests that  the characteristic luminosity
in the quasar luminosity function arises primarily as a consequence of a  
characteristic mass scale above which there is a systematic separation 
between the black hole and the halo merging rates. At lower mass scales, 
black hole merging closely tracks the merging of dark haloes. When combined 
with a declining efficiency of black hole formation with redshift, the 
model can reproduce the quasar luminosity function over a wide range of 
redshifts. The observed space density evolution of quasars is well 
described by formation rates of SMBH above $\sim 10^8 M_\odot$. 
The inferred mass density of SMBH agrees with that found 
independently from estimates of the SMBH mass function derived 
empirically from the quasar luminosity function.

\end{abstract}



\section{Introduction}
Super massive black holes consisting of at least tens of
millions of solar masses are now believed to be lurking in the nuclei of 
most galaxies. These dark objects are also the central engines of the
most powerful sources of light in the universe: the quasars. Discerning 
the evolution history of SMBHs and quasars is a
fundamental challenge for the theories of structure formation in the
universe. Quasars and other forms of Active Galactic Nuclei (AGN) are 
thought to be powered by accretion on to the SMBHs. By virtue of
their enormous brightness these objects are now observable up to
sufficiently high redshifts, corresponding to the dawn of the era of 
galaxy and star formation in the universe. In the context of
hierarchical structure formation, it is believed that mergers of
galaxies play a crucial role in igniting the quasars. The merger induced
gravitational instabilities funnel a huge amount of gas to the 
galactic cores. As a fraction of the kinetic energy of this
gas is released in the form of electromagnetic radiation, luminosities
close to the Eddington luminosity ($\sim 10^{46} \ \rm erg \ s^{-1}$ for
a typical$\sim10^8 M_\odot$ SMBH masses) are produced. 

Nuclear black holes probably trace the merger of their 
host galaxies and coalesce
during mergers. If one envisages a simple scenario where a dark matter
halo harbours a single galaxy and halo merging leads to galaxy merging
along with their SMBHs, a coalescence dominated growth naturally leads
to a linear relation between the black hole and the host halo masses
(see e.g. Haehnelt, Natarajan \& Rees 1998, hereafter HNR98). 
The observed correlation between 
the black hole masses and the bulge velocity dispersion
($M_{BH} - \sigma$ relation) of galaxies \citep{ferraresemerritt}, 
however, suggests that the accreted
component of mass far supersedes the coalesced one. The observed
scaling is concomitant with a scenario where the energy feedback from
the black hole plays a crucial role in regulating the amount of matter
accreted on to it \citep{silkrees}. During the accretion process the 
mass of the SMBH builds up over some characteristic time-scale. 
The ratio of this characteristic time-scale to the Hubble time  
determines the duty cycle of a quasar. Accretion shuts down when the
black hole is so massive that its energy feedback
exceeds the binding energy of its fuel reservoir.
Indeed, as pointed out by HNR98 and \citet{wyitheloeb}, the
observed scaling can be retrieved if the black hole were to supply the
gas with an energy equal to its binding energy over the dynamical time
of galaxies. These intricately woven, yet neat set of ideas sketch an
outline for simple models of quasar and SMBH evolution.

Several analytic and semi-analytic models based on mergers of 
dark haloes have been proposed to explain the high redshift evolution
of quasar luminosity functions (see for e.g. Haehnelt \& Rees 1993,
hereafter HR93; HNR98; Haiman \& Loeb 1998; Kauffmann \& Haehnelt
2000, hereafter KH00; Cavaliere \& Vittorini 2000, hereafter CV00; 
Hosokawa 2002; Wyithe \& Loeb 2003, hereafter WL03; Hatziminaoglou
et al. 2003). Simplified analytic models assume that all the cold
baryons in a halo are locked in a single galaxy and galaxy mergers 
trace the halo mergers. More realistically, the progenitor galaxies 
in a newly formed 
halo sink towards the centre over a dynamical friction time-scale  
(Binney \& Tremaine 1989; Lacey \& Cole 1993) and
merge to form a new baryonic core. However, as the dynamical friction
time-scale is generally much shorter than 
the Hubble time, the approximation that these mergers are
instantaneous, works rather well. Galactic mergers are assumed to 
ignite an AGN 
lasting for a characteristic time
$t_{qso}$. The resulting luminosities are
estimated as Eddington luminosities corresponding to central 
black hole masses, which in turn are determined through their correlation with 
the circular velocities of the newly formed haloes. 

The uncertainty in the mass of cold gas available for
accretion poses a key difficulty for the models. There are two central
aspects of this uncertainty: the first is to ascertain how much cold
gas is available in galaxies of a given mass at a given epoch
\citep{kauffmannhaehnelt} and the second is to understand how
effectively this gas is transported via the galaxy mergers to become 
available for accretion by the forming SMBH \citep{cavalierevittorini}.

The predictions of these models can be tested against the backdrop of a fairly well 
constrained observational data set. The most striking aspect of this data 
is that the quasar population peaks at z $\sim 2.5 $ and exhibits a 
steep decline towards lower redshifts (Schmidt, Schneider \& Gunn
1995; Warren, Hewett \& Osmer 1994; Shaver et al. 1996; Boyle
et al. 2000). At redshifts higher than $ \sim 2.5$ the number 
count appears to decline again, however, the observations at these 
redshift could be marred by
obscuration and extinction effects. Simple analytical models mentioned
above can well explain the high redshift observed luminosity functions ($z \gtrsim
2$). They, however, systematically fail at lower redshifts ($z
\lesssim 2$) and this epoch in the history of the universe is
somewhat opaque in our understanding of the evolution of
quasars. Even though the works of CV00, HR93 and KH00 shed important light in this 
regard, no comprehensive scheme is yet available to  systematically 
quantify the various aspects of quasar evolution at $z \lesssim 2$.

The present work is primarily motivated by the low redshift anomaly
between observations and theoretical predictions. To bridge this gap
we introduce a toy model centred around two main ideas: the first pertains 
to the effect of declining black formation efficiency at low
redshifts and the second involves a declining black hole formation
rate in haloes above some characteristic mass $M_*$. The first idea is
derived from a model investigated in \citet{haehneltrees} where it is
argued that the conditions for SMBH formation are more favourable at higher 
redshifts as compared to low redshifts. The second idea is based on
the existence of a break luminosity in the empirical quasar
luminosity functions. 
We attribute this break luminosity to
a systematic decoupling between halo and galaxy merger
rates in haloes above a characteristic mass along the line suggested by CV00 that galaxy
merging becomes rarer in sufficiently massive haloes, where individual
galaxies retain their properties. Furthermore, in a similar context, 
\citet{haimanciottiostriker} have argued that a decrease in the quasar
luminosity density towards low redshifts arises due to a decline in the 
formation rates of spheroids. Here we propose a characteristic mass
scale $M_*$ for haloes which marks a 
drop in the formation rates of black holes and thus possibly spheroids 
as well. As the low redshift universe is an era of group assemblage, our  
value of $M_*$ which corresponds to a group mass scale seems
natural to explain the steep decline observed in the quasar luminosity density
at low $z$.

Our model entails a phenomenological treatment of the ideas
described above. Notwithstanding the crudeness of the approach, 
it yields interesting results for the the quasar luminosity 
functions and their evolving number density, broadly consistent with
the observations. It also provides a simple explanation for the 
evolution of the characteristic luminosity of quasars. 
Furthermore, the minimum SMBH masses for quasars suggested by the
model (by comparison with the observed space density evolution of
quasars) are $\gtrsim 10^8 M_\odot$, which is consistent with the
original estimates of \citet{soltan}. Finally the model can be linked quite 
simply with the evolution of cold gas masses in
galaxies and star formation rates in the universe and thus yields estimates 
of the contributions to the ionising background at high redshifts,
due to star formation and quasar activity.

The structure of the paper is 
as follows: in section 2 we review some important aspects of analytical 
modelling of hierarchical structure formation which provides the
essential tools for our model; 
section 3 discusses the relation between the black hole masses and
their host haloes in a scenario with a declining efficiency of SMBH
formation with redshift. Section 4 extends the argument of major
mergers of haloes to compute the SMBH formation rates and the
luminosity functions of the ensuing quasars. In section 5 we compare
the SMBH formation rates from our model against those inferred from
the empirical double power-law fit to the quasar luminosity
functions. In section 6 we estimate the mass functions of SMBH and the
comoving cosmological mass density in the SMBH at various
redshifts. Section 7 describes how to obtain simple estimates of the cold gas
density and star formation rate density in the universe; section 8
extrapolates these star formation rates to high redshift to compute
the ionising background. These are then
compared with a similar extrapolation to high redshifts, of the quasar 
luminosity density. We summarise and discuss our results
in section 9.

\section{Elements from extended Press-Schechter theory}
Our approach largely takes advantage of the extended
Press-Schechter (EPS) formalism which provides essential tools 
for computing the merger probabilities and rates of bound objects 
(Bond et al. 1991; Bower 1991; Lacey \& Cole 1993, hereafter LC93). 
In this scenario the initial density fluctuations 
in the universe are assumed to be Gaussian in nature. 
The growth of these fluctuations proceeds linearly
initially and is analysed through the linearised ideal fluid equations
(e.g. Silk \& Bouwens 2001). The fluctuations subsequently turn non-linear and
eventually collapse to form bound objects as the density reaches a critical
threshold. A comparison of EPS results with the results of N-body
simulations as given in \citet{laceycole2}, demonstrates that 
the prescription works rather well.

The linear growth of fluctuations is described through the linear
growth factor $D(t)$. For a critical cosmology this factor is
simply equal to the scale factor $a$ for that epoch. For the $\rm \Lambda CDM$
cosmology used in this paper ($\Omega_{m}^0 = 0.3, \
\Omega_{\Lambda}^0 = 0.7,\ \Omega_{b}^0 = 0.02/h^2, \ \sigma_{8} =
0.9, h = 0.7$), we compute it as given in \citet{nfw}. 
The mass M of an inhomogeneity is represented through a sharp k-space 
filter which is used to smooth the
matter density field at this mass scale. As suggested in LC93, $M$ can be related to 
$k_s$ through a qualitative mapping $M=6\pi^2 \rho_0 k_s^{-3}$. Here $\rho_0$ is
the mean background density of matter in the universe. Using the dark matter
power-spectrum $P(k)$ parametrized as in \citet{ebw}, the variance
for mass scale M can then be simply calculated as
\begin{equation}
\sigma^2 \ (M) \  =  \int_0^{k_s} \ P(k) \ 4 \pi \ k^2 \ d k.            
\label{sigma}
\end{equation}
The mass function of haloes is now written in terms of $\sigma^2 (M)$ as
\begin{equation}
N_{PS} (M,t)  = \sqrt{ \frac {2}{\pi}}
\frac {\bar {\rho_0}}{M} \frac {\omega}{D(t)} \left|\frac{1}{\sigma^{2}}
\frac {\d\ \sigma}{\d  M} \right |  \rm{exp} \left [-\frac {1}{2}\frac
{\omega^2}{D^2(t)\sigma^2} \right ] ,
\label{nps}
\end{equation}
where $\omega$ is the critical density for collapse. 

The formation rates, major merger rates and survival probabilities have 
been discussed by several authors (e.g. Sasaki 1994; Kitayama \& Suto
1996, hereafter KS96). In this context, the excursion set approach of
\citet{bondetal} (also LC93) yields two important conditional probabilities. 
The first is the conditional
probability that a halo of given mass $M_2$ at time $t_2$, had a progenitor
of mass $M_1$ at earlier time $t_1$. This is given as
\begin{eqnarray}
P_{1}(S_1,\omega_1|S_2,\omega_2) \d S_1 & = & \frac{1}{\sqrt{2\pi}} \frac{(\omega_{1} -
\omega_2)}{(S_1-S_2)^{3/2}} \nonumber \\
& & \times \rm {exp} \left [ - \frac{(\omega_1 -
\omega_2)^2}{2(S_1-S_2)} \right ] \d S_1
\label{p1},
\end{eqnarray}
where $\omega_1$ is the critical density at time $t_1$ and $\omega_2$
at time $t_2$. Also, $S_1$ is the variance corresponding to mass $M_1$
and $S_2$ is the variance corresponding to $M_2$.
The relation between the critical density $\omega(t)$
at any time $t$ is given as $\omega(t) = \omega_0/D(t)$, where
$\omega_0$ is the present day critical density ($\sim 1.686$ for
a SCDM cosmology). 

The second is the probability that a halo of mass $M_1$ at
time $t_1$, will result in a larger halo of mass $M_2$ at later time
$t_2$ can be written as
\begin{eqnarray}
P_{2}(S_2,\omega_2|S_1,\omega_1) \d S_2 & = & \frac{1}{\sqrt{2\pi}} \frac{\omega_2(\omega_1 -
\omega_2)}{\omega_1}\left [ \frac{S_1}{S_2(S_1-S_2)} \right ]^{3/2}
\nonumber \\
& & \times \rm{exp} \left [ - \frac{(S_2\omega_1 -
S_1\omega_2)^2}{2 S_1 S_2(S_1-S_2)} \right ] \d S_2.
\label{p2}
\end{eqnarray}
Through appropriate transformations the conditional
probabilities can also be written in terms of halo masses. 

\subsection{Merger rates, survival probabilities and formation epoch distributions}

The EPS formalism lacks a self sufficient way to compute the formation
rates of haloes. This is because effectively all haloes  
at any epoch are continuously accreting mass and in this sense are
newly created. There is no clear distinction between
the objects accreting less mass and the objects
which accrete substantial mass. For computing the major merger rates,
therefore, a criterion has to be introduced by hand. 
One has to assume that if none of the
progenitors of a newly formed object had mass greater than half (or
some other fraction) of the object mass, the event is a major
merger. Alternatively Sasaki (1994) proposed to derive the formation rates by 
assuming that the destruction probability of objects (i.e. the
probability that they get incorporated into higher mass objects) has no 
characteristic mass scale. This rate is then simply the positive term in
the time derivative of the PS mass function. In Sasaki's prescription 
the survival probability of haloes from one redshift to another is
simply the ratio of the growth factors at these redshift. The product 
of formation rate and the survival probability defines the formation epoch
distribution. Interestingly for a critical cosmology, Sasaki's
formation epoch distribution can be integrated analytically over all
redshifts to yield the exact mass function of haloes.

In the present work we compute the major merger rates based on the
prescriptions described in LC93 and KS96.The probability that a 
given halo of mass $M_2$ at time
$t_2$ had, at some earlier time $t_1$, a progenitor with mass below some 
threshold mass $M_c (<M_2)$ is computed by integrating equation \ref{p1} over $S_1$
from $S(M_c)$ to $\infty$. If the difference between the times $t_1$
and $t_2$ is substantial, then $M_1$ can reach the value $M_2$ through 
either a single merger occurring
anytime between $t_1$ and $t_2$ or it can do so through more than one
merger/accretion episodes between the two times. However, if the
difference between the two times is very small then the entire change
in mass ($\Delta M = M_2 - M_1$) can be treated as arising from a
single merging event. Letting $\Delta \omega = \omega_1 - \omega_2$ we 
thus write 
\begin{eqnarray}
P(M_1 < M_c, t_1|M_2,t_2) &=&  \int_{0}^{M_c} P_1 \ \rm{d} \it
M_1 
\label{pint} \\ \nonumber
&=&  \rm{erf} \it \it\left[ \frac{\Delta \omega}{\sqrt{\rm{2} \it
      (S_c - S_{\rm{2}})}} \right],
\end{eqnarray}
where $S_c$ is the variance corresponding to mass $M_c$. 
For small $\Delta \omega$ taking the series expansion in equation 
\ref{pint} up to first order only, we compute the major merger
rates of dark haloes as 
\begin{equation}
R_{MM}(M,t)  = \sqrt{\frac{2}{\pi}}  \frac{1}{\sqrt{ S(M/2)
    -S(M)}}\times \Bigg | \frac{ d  \omega}{d t} \Bigg | N_{PS}(M,t),
\label{mm}
\end{equation}
where we have chosen $M_c = M/2$. 
Note that when $S_c$ tends to $S$, the right hand side of equation
\ref{pint} tends to one, if $\Delta \omega$ is small but finite. 
This implies that at any given epoch, all the existing objects 
continuously accrete mass. 
\begin{figure}
\begin{center}
\includegraphics[width=85mm]{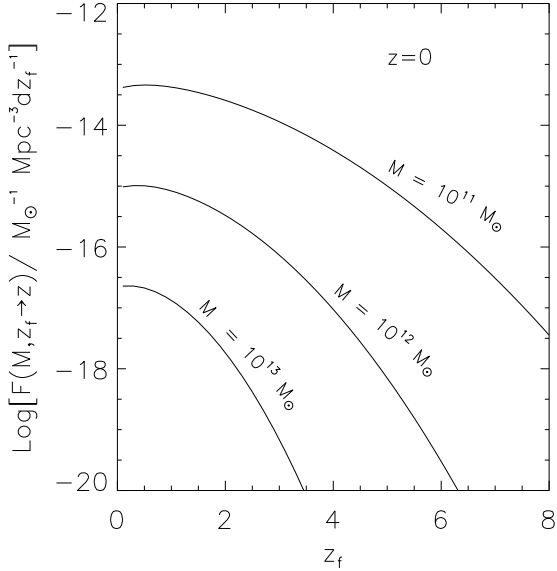}       
  \caption{The formation epoch distribution of DM haloes of different masses as
  given by equation \ref{fhalo}.}
\label{formepoch}
\end{center}
\end{figure}

As pointed out in LC93, when the halo mass for a
trajectory falls to a smaller value at an earlier time, it only
means that {\em one} of the parent haloes had this smaller mass. 
As a result, it is always
possible that the largest coeval progenitor already has more than half the
mass of the final halo at this epoch. Such an event cannot be characterised as a major merger as 
we define it. The approach of KS does not account for this
possibility. Although this marginally alters the rates for the
range of halo masses we are concerned with, it is nevertheless
interesting to see how one can systematically account for this
possibility and therefore, we propose a solution to this problem in appendix A.

Equation \ref{p2} can be used to calculate the survival probabilities 
as discussed in LC93 and KS96.  These are based on the assumption that
an object retains its identity up to the time it doubles its mass. The
probability that a halo of mass $M$ at present time $t_1$, will not 
undergo such a merger/accretion, before time $t_2$, which would result in 
its final mass being greater than $2M$, is given as
\begin{equation}
P_{surv}(M,t_{1},t_{2}) =  \int_{M}^{2M} P_{2}(M_{2},t_2|M,t_1) \d M_{2}.
\label{psurv}
\end{equation}

\begin{figure}
\begin{center}
\includegraphics[width=85mm]{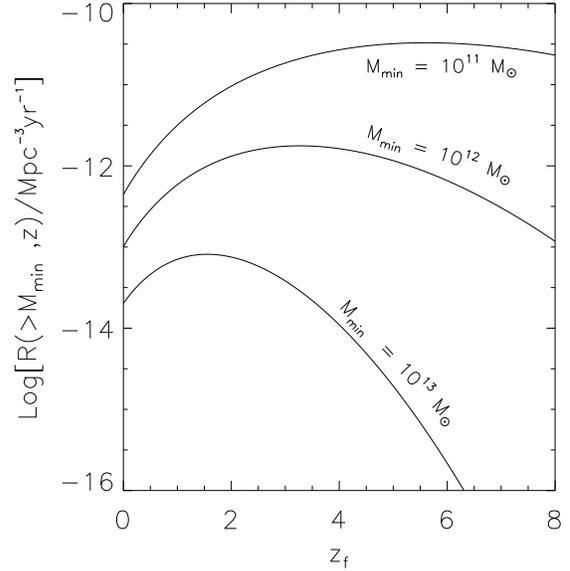}       
  \caption{ The evolution of cumulative formation rates of dark matter haloes with
  masses above $M_{min}$. These rates (particularly for $M_{min} =
  10^{12} M_\odot$, the typical quasar host halo mass) do not produce as steep
  a decline at low redshifts as that measured for quasar number counts.}
\label{cumformrate}
\end{center}
\end{figure}

As discussed previously, the survival probability of haloes multiplied
with their merger rates, yields their formation epoch
distribution. This distribution tells us that in a given population of 
mass $M$ objects at time $t$, what fraction were 
formed at an earlier time $t_f$ and have retained their identity since 
then. Thus the formation epoch distribution in terms of formation time 
$t_f$ can be written as
\begin{eqnarray}
F(M,t_f \rightarrow t) \d M \d t_{f} &=& R_{MM} (M,t_{f}) \times
P_{surv} (M,t_{f} \rightarrow t) \nonumber \\ 
&& \ \times \ \d M  \ \d t_{f}.
\label{fhalo}
\end{eqnarray}
We show in Fig~\ref{formepoch} the formation epoch distribution of haloes of
different masses as a function of redshift, and in Fig~\ref{cumformrate}
the formation rates of objects with masses above 
$10^{11}$, $10^{12}$ and $10^{13} \ M_\odot$  respectively. 
It can be seen from Fig~\ref{cumformrate} that merger rates of haloes
alone do not produce a significant evolution in the number
density at low $z$, and therefore a naive model directly linking quasar triggering
to merger rates of haloes would fail to match the steep 
decline which is observed in their number density. Hence, we subsequently discuss
the additional inputs that need to be incorporated into the model
in order to reproduce the observed evolution of the quasar population.

\subsection{Halo circular velocities}

For a nearly isothermal halo, the mapping between the halo circular
velocity and its mass can be obtained using the relations given in
\citet{somervilleprimack}
\begin{equation}
  v_{c}^2  = \frac{GM}{r_{vir}(z)} -
\frac{\Omega_{\Lambda}}{3}H_0^{2} \ r_{vir}^2 (z),
\label{vcm}
\end{equation}
where $z$ is the collapse redshift and $r_{vir}$ is the virial
radius. $r_{vir}$ can be computed by noting that 
$M = \ 4 \pi/3 \ r_{vir}^3(z) \ \rho_{vir}(z)$ where the 
virial density $\rho_{vir}(z)$ is given as
\begin{equation}
\rho_{vir}(z) = \Delta_c(z) \rho_c^{0} (1+z)^3 
\frac{\Omega_0}{\Omega(z)}.
\label{rhovir}
\end{equation}
Here, $\Omega_0$ and $\Omega(z)$ are the matter densities of the
present day and redshift $z$ universe respectively; $\rho_c^0 = \ 2.8
\times 10^{11} \ h^2 \ \rm{} M_\odot Mpc^{-3} \it{}$ is the
present critical density of the universe and a fitting
expression for $\Delta_c(z)$ is given as $\Delta_c(z) = 18 \pi^2 + 82
x - 38 x^2$, where $x = \Omega_m(z) -1 $ \citep{bryannorman}. 
Thus the expression for virial radius in terms of formation redshift
$z$ and halo mass $M$ is given as
\begin{equation}
r_{vir} = 2.14 \times 10^{-5} M^{1/3}
\left[\frac{\Omega_m^0}{\Omega_m(z)}\frac{\Delta_c(z)}{18\pi^2}\right]^{-1/3}
(1+z)^{-1} \ \rm{Mpc}.
\label{rvir}
\end{equation}
These relations together enable us to write the halo circular 
velocity as
\begin{equation}
v_c = 1.38 \times 10^{-2}
M^{1/3}\left[\frac{\Omega_m^0}{\Omega_m(z)}\frac{\Delta_c(z)}{18
  \pi^2}\right]^{1/6} (1+z)^{1/2} \ \rm{km s^{-1}} \ .
\label{vc}
\end{equation}

\section{Super Massive Black Hole masses}

In a self-regulated growth of SMBH masses, the correlation between
the galaxy circular velocities ($v_c$) and SMBH masses ($M_{BH}$) is 
written as \citep{silkrees}
\begin{figure}
\hskip 5pt
\epsfxsize 8.5 cm
\epsfbox{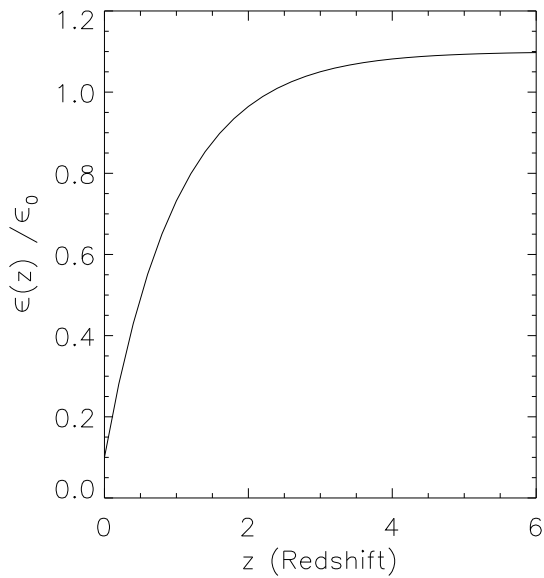}
\caption{Phenomenological redshift evolution of the formation
  efficiency of black holes (see text for the exact function).
  The characteristic redshift of decline is
  $z_* = 1$. A steep decline is obtained around $z = 0$ where values become 
as low as ten percent of the maximum value.}
\label{efficiency}
\end{figure}

\begin{equation}
M_{BH} \propto  v_c^5,
\label{mbhvc}
\end{equation}
where the constant of proportionality depends on the fraction of cold
gas $f_{cold}$ available for accretion by the coalesced seed black hole and weakly 
on the presence of an accretion disc (HNR98; Burkert \& Silk 2001). Such a
scaling is corroborated by the investigations of
\citet{ferraresemerritt} and \citet{gebhardtetal}.

Incorporating $f_{cold}$ in an analytic model is not simple. 
Firstly, we do not know how much cold gas exists in haloes of
different masses at different epochs and secondly, even if we knew the
mass of cold gas, it is impossible to analytically compute the
contributions coming from different mass progenitors during a
merger. For the sake of simplicity, therefore, we get around this 
problem by adopting an empirically motivated decline in black hole
formation efficiency with redshift. This can be attributed to a 
general decline in the amount of cold gas in the galaxies at low 
redshifts. Such an approach has been considered previously, for example, by 
\citet{haehneltrees}. 

We model the declining formation efficiency of SMBH masses through a function
$\epsilon (z)$ such that $M_{BH} = \epsilon(z) \ v_c^5$ and consider a
phenomenological form of $\epsilon (z)$ as
\begin{equation}
\epsilon (z) = \epsilon_0 \times \left[ c  - \rm exp \it \left( -
  \frac{z}{z_*} \right) \right].
\label{effz}
\end{equation}
Here $\epsilon_0$ and $c$ are constants and $z_*$ is a characteristic
redshift. We choose a value $z_* \approx 1$ based on the observed
peak in the star formation rate (SFR) density in the universe at 
redshifts between $1$ and $2$ (see e.g. Ascasibar et al. 2002; 
Silk \& Rees 1998). This is because a
peak in the SFR density is a general indicator of a gas rich environment
in galaxies. It is, however, still a matter of debate whether SFR
density peaks at $z \sim 1 - 2$ or reaches a broad maximum there. To
obtain a steep decline we choose $c = 1.1$ such that the minimum value
of $\epsilon (z)$ is roughly 10 percent of the maximum value $1.1 \times
\epsilon_0$. This latter is chosen to match the
observed $M_{BH}-\sigma$ correlation \citep{ferrarese}
\begin{equation}
M_{BH} = 1.66 \times 10^8 \left( \ \frac{\sigma}{200 \ \rm km 
\ s^{-1} \it}  \ \right)^\alpha M_{\odot},
\label{mbhvc}
\end{equation}
where we adopt $\alpha = 5$ consistent with a feedback 
regulated growth of SMBH masses. This is slightly different from the
value $\alpha = 4.58 \pm 0.52$ inferred from SMBH mass measurements,
the difference possibly having its origin in the observed
non-linearity of $v_c - \sigma$ relation (Ferrarese 2002; WL03).
For simplicity, we consider the linearly fitted relation for ellipticals 
$\sigma \approx 0.65\ v_c$ to obtain $\epsilon_0 = 6 \times 10^{-5}$
(such that $M_{BH} \sim \epsilon_0 \ v_c^5$ at high redshifts).

Figure \ref{efficiency} shows a plot of efficiency versus the
formation redshift, i.e. how equation~\ref{effz} behaves. We can now write the black hole masses in 
terms of formation redshift $z$ and halo masses $M$ as
\begin{eqnarray}
M_{BH} &=& \epsilon (z) \ \times 5 \ \times 10 ^{-10} \left(
\frac{M}{M_{\odot}} \right)^{5/3} \nonumber \\
&& \hskip 0.5 cm \times \ \left[\frac{\Omega_m^0}{\Omega_m(z)}
\frac{\Delta_c(z)}{18 \pi^2}\right]^{5/6} (1+z)^{5/2} \ M_{\odot}.
\label{mbhz}
\end{eqnarray}

\section{Formation rates of SMBH and quasar luminosity functions}

We now use the $M_{BH} - M_{halo}$ mapping (equation \ref{mbhz}) to 
transform the halo major merger rates into the black hole formation
rates in terms of the formation redshift $z$ as 
\begin{equation}
R_{BH}^1 (M_{BH},z) = R_{MM}[M_{halo}(M_{BH},z),z] \frac{d M}{d
  M_{BH}} (M_{BH},z). 
\label{rbh1} 
\end{equation}
The superscript $1$ in $R_{BH}^1$ is 
to differentiate it from a modified rate 
$R_{BH}$, which takes into account a decline in black
hole formation rates in haloes above a characteristic 
mass $M_*$. This is achieved through a phenomenological 
function $F_{\mu}(M_{halo})$ such that
\begin{equation}
R_{BH} ( M_{BH},z) = R_{BH}^1 ( M_{BH},z) \times
F_{\mu}[M_{halo}(M_{BH},z)].
\label{rbh2} 
\end{equation}
\begin{figure}
\hskip 5pt
\epsfxsize 8.5 cm
\epsfbox{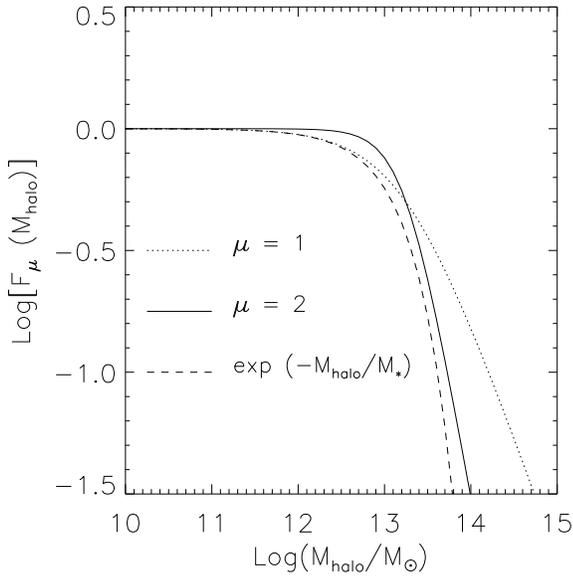}
\caption{This figure shows the second of our two phenomenological
functions. We use this function to introduce a drop in the formation
rates of black hole in haloes above the characteristic mass scale
$M_*$, the value of which has been set by comparison with the quasar
luminosity function (figure \ref{lf})to about $M_* =
10^{13.25}M_\odot$. The redshift evolution
of the luminosity corresponding to this characteristic mass
is plotted against the empirical characteristic luminosity of the
double power law fit, in fig \ref{critlum}.}
\label{mu}
\end{figure}
For the form of $F_{\mu}(M_{halo})$, we found
that a simple exponential function $\rm exp \it (-M/M_*)$,
with a cutoff at $M_*$ is too steep to be 
consistent with the observed luminosity functions. We therefore adopt a 
function with a following behaviour:
$F(M) \rightarrow 1$ for $M << M_*$ and $F(M) \rightarrow
(M_*/M)^{\mu}$ for $M >> M_*$, where $\mu$ is a positive number. 
Instead of a sharp exponential cut-off, such a function has 
softer `power-law' drop at the high mass end: 
\begin{equation}
F_{\mu}(M) = \frac{1}{1 + (M/M_*)^\mu}.
\label{functionm}
\end{equation}
$F_\mu(M)$ for values of $\mu = 1$ and $\mu =2$ is plotted in figure
\ref{mu} along with the exponential 
function $F(M) = \rm exp\it(-M/M_*)$.

It is important to mention at this point that in massive haloes ($M > M_*$), 
rarer galaxy mergers could possibly alter the $M_{BH} - M_{halo}$
mapping. Furthermore, the AGN could enter the 
sub-Eddington regime due to lack of gas mass transported through rare 
galaxy mergers. However, to keep our model simple we assume that even 
for ineffective 
galaxy mergers, the $M_{BH}-M_{halo}$ scaling is as given in
equation \ref{mbhz}.
\begin{figure}
\hskip 5pt
\epsfxsize 8.5 cm
\epsfbox{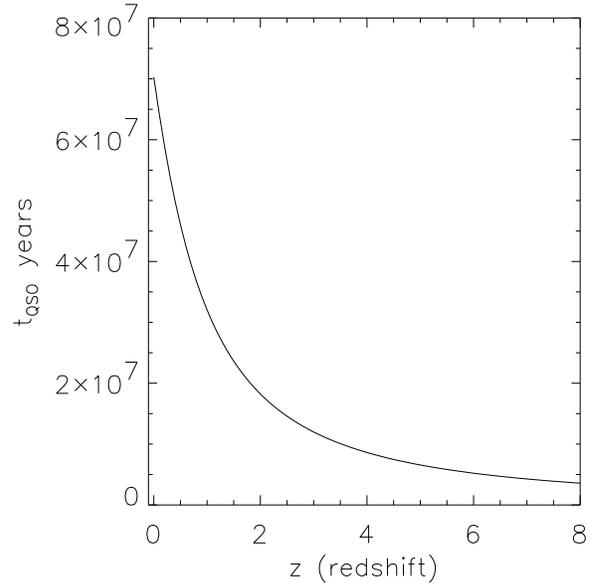}
\caption{The model lifetime of quasars given as $t_{qso} = k \ t_{dyn}$,
  where $t_{dyn}$ is the halo dynamical time and the constant of
  proportionality is about $k \approx 0.035$ (e.g. WL03)}.
\label{tqso}
\end{figure}
If a merger-ignited black hole shines at the Eddington 
luminosity $L_{bol}/L_{Edd} = 1$, we can write the $B-$band
luminosities of AGNs as
\begin{equation}
L_B = f_B \times 1.3 \times 10^{39} \left[ \frac{M_{BH}}{10^8
    M_\odot} \right]\rm{} J \ s^{-1}, 
\label{lb} 
\end{equation}
where $f_B$ is the fraction of bolometric luminosity emitted in
the B-band. Realistically, as the accretion proceeds, the black hole
mass builds up over the characteristic accretion time. This process is
accompanied by a parallel increase in the quasar
luminosity which determines its light curve. However, a
simplification is possible if one considers that the black hole mass
instantaneously increases to post merger values given by equation
\ref{mbhz}. The black hole then shines at the Eddington
luminosity, corresponding to this new mass, over characteristic
time-scale $t_{qso}$ which lies in the range  
$t_{qso}/t_{Hubble} \sim 10^{-2} \ \rm{to} \ 10^{-3}$. Thus using 
the $L_B - M_{BH}$ mapping and the fact that $t_{qso} \ll t_{Hubble} $
we can simply write the quasar luminosity function as 
\begin{equation}
\Phi(L_B,z) \simeq t_{qso} \times R_{BH} [M_{BH}(L_B),z] \times \frac{d
  M_{BH}}{d L_B } (L_B,z). 
\label{lf}
\end{equation}
\begin{figure*}
\hbox{\epsfxsize 18 cm
\epsfbox{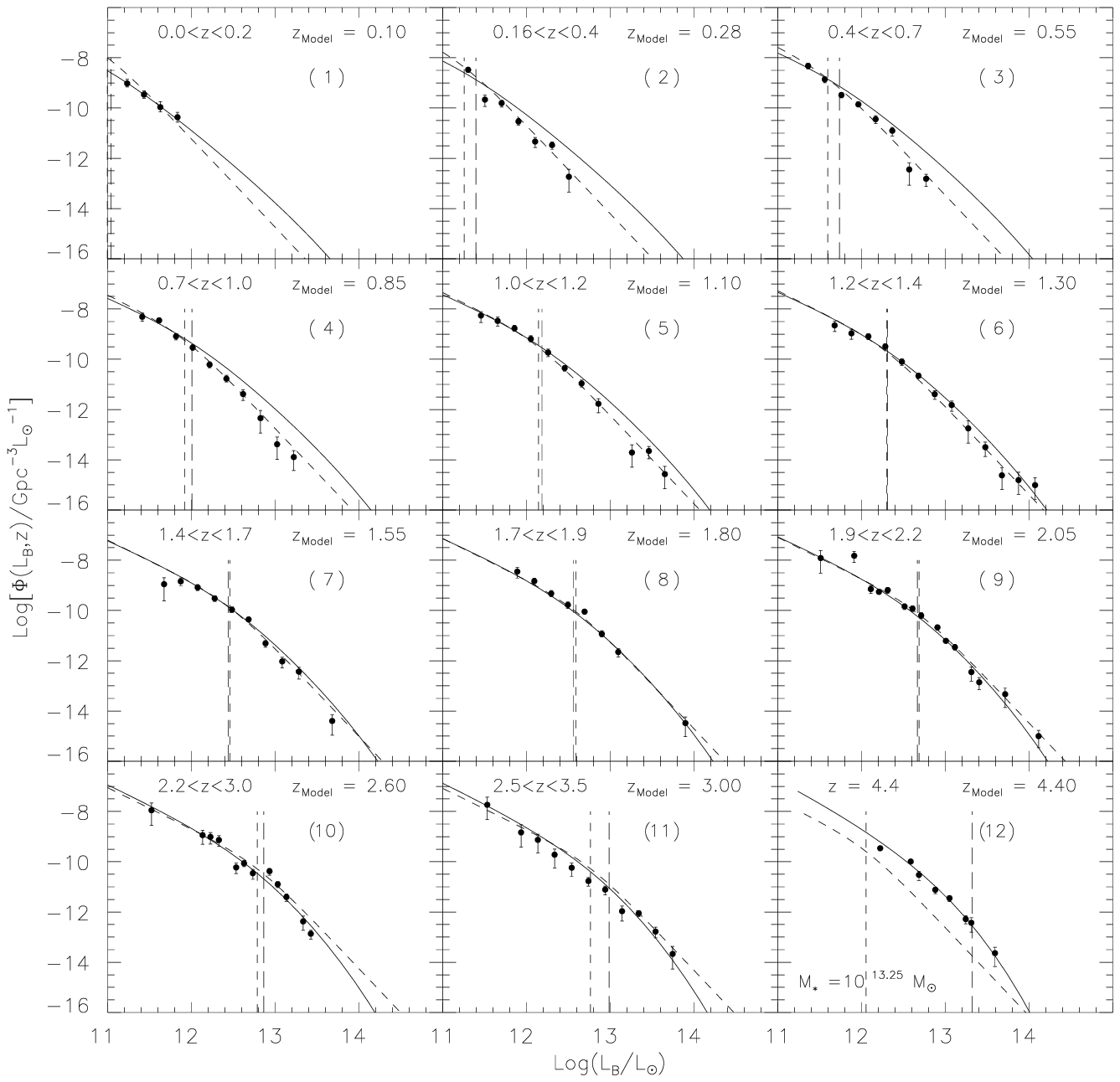}}
\caption{Model luminosity functions (solid curve) plotted against the observed
  luminosity functions (filled circle) of quasars. The data for $z
  \lesssim 4$ is from \citet{pei} and for $z \approx 4.4$ is from 
  \citet{kennefick}. The model parameters are
  $\epsilon_0 = 6 \times 10^{-6}$, $\mu = \rm 2$, $M_* = 10^{13.25}
  M_\odot$ and $t_{qso} \sim 0.035 \ t_{dyn}$. The short-dashed curves 
  show the double power
  law fit of \citet{pei}. The long-dashed vertical line is the model
  break luminosity corresponding to $M_*$ and the
  short-dashed vertical line is the characteristic luminosity in the
  double power-law fit.}
\label{luminosityfunction}
\end{figure*}

The quasar lifetime, $t_{qso}$ is not a very well constrained quantity. Two
relevant time-scales for quasars are the Salpeter (or 
e-folding) time-scale and the dynamical time-scale of galaxies. The Salpeter
time-scale is given by 
$t_{Sal} = 4 \times 10^7 (\epsilon_{rad}/0.1)$ where $\epsilon_{rad}$ 
is the radiative efficiency of black holes
defined through the relation $L_{bol} = \epsilon_{rad} \dot{M}
c^2$, $c$ is the speed of light and $\dot{M}$ is the accretion rate
of matter on to the black hole.
On the other hand it has been noted, for example, in HNR98 and WL03,
that the observed $M_{BH}-v_c$ relation naturally arises if the black hole
injects the gas with an energy equal to its binding energy, over a
local dynamical time-scale of the galaxy. In this respect, the local dynamical
time of a galaxy is also a natural time-scale. In the present model, 
following WL03, we adopt a quasar lifetime proportional to the local 
dynamical time of 
galaxies in newly formed haloes. The local dynamical time of galaxies being 
themselves proportional to the halo dynamical time, we get $t_{qso} \propto
t_{dyn}$, where $t_{dyn}$ is the halo dynamical time and so we write
\begin{figure}
\hskip 5pt
\epsfxsize 8 cm
\epsfbox{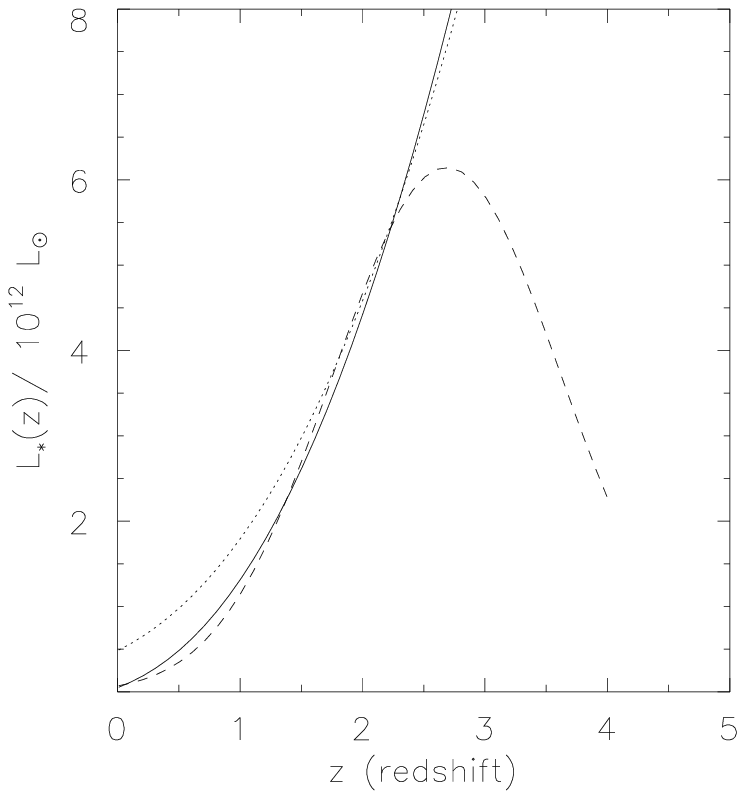}
\caption{The evolution of break luminosity in our model
  (solid line) compared with the characteristic luminosity 
  in the double power law
  luminosity function fit (dashed line). Also plotted is the
  break luminosity in a scenario where $\epsilon =
  \epsilon_0$ is fixed (dotted line). The characteristic 
  mass is $M_* = 10^{13.25} M_{\odot}$. }
\label{critlum}
\end{figure}
\begin{figure}
\hbox{\epsfxsize 8.5 cm
\epsfbox{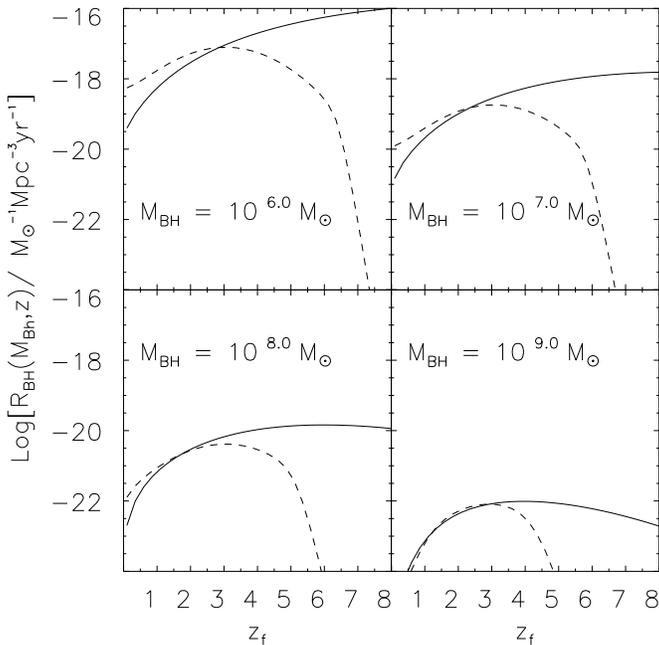}}
\caption{This figure shows the redshift evolution of formation rates
  of black holes with masses $10^6$,$10^7$,$10^8$ and $10^9 \ M_\odot$
  respectively. The dashed lines are from equation \ref{rbhobs}
  and the solid line are from equation \ref{rbh2}}
\label{lisa}
\end{figure}
\begin{figure}
\hbox{\epsfxsize 8.5 cm
\epsfbox{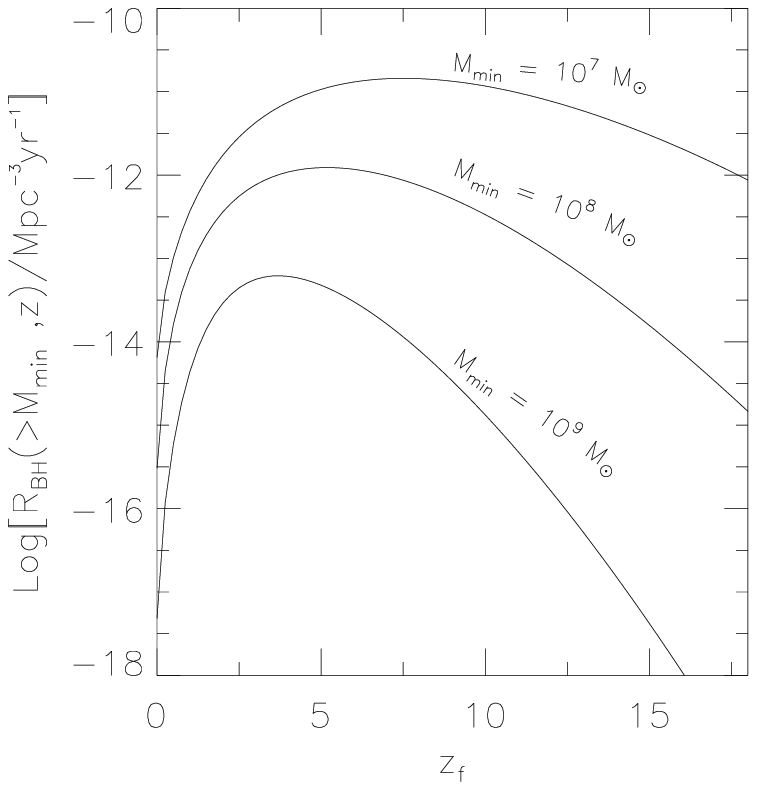}}
\caption{Cumulative formation rates of black holes with masses above
  $10^7$, $10^8$ and $10^9 \ M_{\odot}$ respectively.}
\label{cumbhrate}
\end{figure}
\begin{equation}
t_{qso} = k \times 1.5 \times 10^9
\left[\frac{\Omega_m^0}{\Omega_m(z)}\frac{\Delta_c(z)}{18
    \pi^2}\right]^{-1/2} (1+z)^{-3/2}\ \rm{yr},
\label{tqsotext}
\end{equation}    
where the constant of proportionality is $k\sim 0.035$ (e.g. WL03). 

In figure \ref{luminosityfunction} we plot the luminosity 
functions as given by equation \ref{lf} (solid curves). The parameters are $\mu = 2$, 
$M_* = 10^{13.25} \ M_\odot$ and $f_B = 0.06$.  
The short-dashed curve is the double power law fit as given
in \citet{pei}. We transform the data and the double power-law fit to
a $\Lambda$ cosmology by noting that $\Phi(L,z) \propto dV^{-1}
d_L^{-2}$ and $L \propto d_L^{2}$, where $dV$ is the volume element
and $d_L$ is the luminosity distance (e.g. Hosokawa 2002). The double
power-law is given as
\begin{equation}
\Phi_{obs}(L_B,z) =  \frac{\Phi_*/L_z}{(L_B/L_z)^{\beta_l} + (L_B/L_z)^{\beta_h}}, 
\label{dpl}
\end{equation}
where $\rm log(\Phi_*/\rm Gpc^{-3}) = 2.95 \pm 0.15$, $\beta_l = 1.64 \pm
0.18$ and $\beta_h = 3.52 \pm 0.11$. The characteristic luminosity
$L_z$ has a Gaussian form 
\begin{equation}
L_z = L_* (1+z)^{-(1+\alpha)} \rm \exp \it \left[ - \frac{(z -
    z_*)^{\rm 2}}{\rm 2 \it \sigma_*^2 } \right],
\label{cl}
\end{equation}
where $\rm log \it (L_*/L_\odot) \rm = 13.03 \pm 0.10$, $z_* = 2.75 \pm
0.05$, $\alpha = -0.5$ and $\sigma_* = 0.93 \pm 0.03$. In figure~\ref{luminosityfunction}, the vertical
short-dashed line is the characteristic luminosity given by equation
\ref{cl} and the vertical long-dashed line shows the break
luminosity for our model, corresponding to $M_* = 10^{13.25}M_\odot$
haloes. It can be seen from the figure that the model can
systematically account for the observed data over a wide range of
redshifts. At $z < 1$, where the break luminosity of the model is
slightly greater than the empirically fitted characteristic
luminosity, we find that the bright end is slightly over predicted. 
The characteristic luminosity and the break luminosity 
follow each other closely up to $z\approx 2.5$.
This is especially clear in figure~\ref{critlum}, where we additionally 
show (dotted curve) the break luminosity corresponding to $M_*$ 
in a fixed black hole formation efficiency scenario. 

It has been suggested in WL03 that rather than being an intrinsic
feature of the quasars themselves, the break in the luminosity
function is present only at low redshifts. Our model suggests otherwise. 
We find that the break in the luminosity function exists as a
consequence of the characteristic mass and is present over a wide
range of redshifts. Furthermore, the double power law in the form given above 
under-predicts the quasar luminosity function at $z = 4.4$. 
On the other hand, inserting the break luminosity given by our
model into the double power law description of the data over-predicts 
the quasar luminosity function. A possible implication of 
this is that perhaps the black hole formation efficiency declines 
beyond $z \sim 4$.  

The tuning of parameters for our model luminosity function is admittedly
crude. Also, it can be seen from equations
\ref{mbhz} and \ref{lb} that there exists a degeneracy between the parameters
$\epsilon_0$ and $f_B$. Higher values of $\epsilon_0$ require
systematically lower values of $f_B$. 
Nevertheless our fiducial numbers are in broad agreement with the values
obtained by other investigations, and we emphasize once again that the 
model presented here only attempts to discern the broad trends of 
SMBH and quasar population evolution.
\begin{figure*}
\hskip 10 cm
\hbox{\epsfxsize 12 cm
\epsfbox{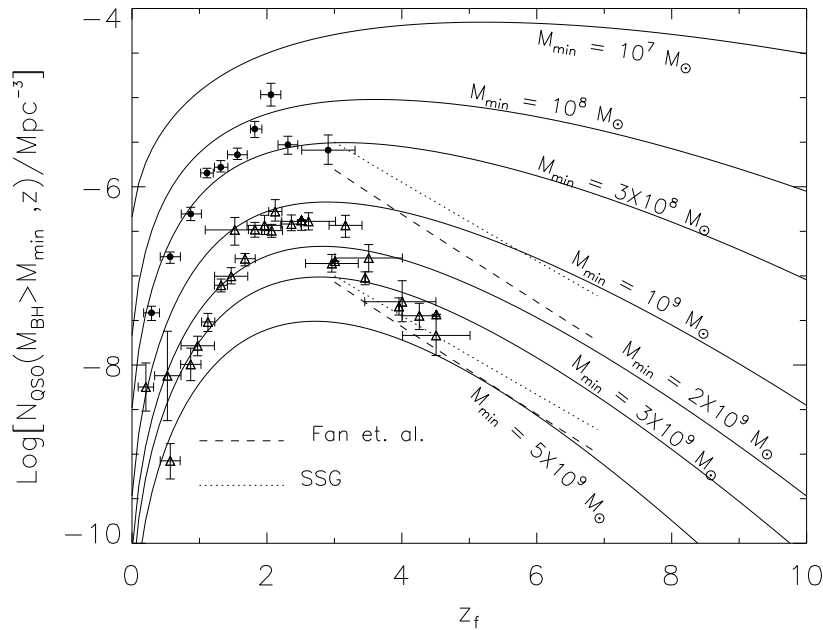}}
\caption{ The space density evolution of quasars with minimum black hole mass
  $M_{min}$ in the model, plotted against the
  observational data. Filled circles describe the observed space
  density evolution of quasars with absolute B-band magnitudes
  $M_B < -24$ and open triangles describe the observed space density
  evolution of quasars with absolute B-band magnitudes $M_B <
  -26$. The data points are plotted as in Kauffmann \& Haehnelt 2000
  (see references therein). Also shown are the best fit to the 
  high redshift data from \citet{fanetal} (dashed lines) and
  \citet{schmidtschneidergunn} (dotted lines). The model parameters are same as in figure 
  \ref{luminosityfunction}.}
\label{NQSO}
\end{figure*}

\section{Formation rates of SMBH from the observed luminosity functions}

Due to their emission of gravitational waves, the merger rates of SMBHs
will be detected directly by instruments such as the Laser 
Interferometer Space Antenna (LISA). In our model, SMBH
merger/formation rates are given by equation \ref{rbh2}. 
One can also use the double power-law
fit to the luminosity function data to estimate these rates 
as a function of SMBH mass. These can be written as
\begin{eqnarray}
R_{BH}^{obs}(M_{BH},z) &=& \frac{1}{t_{qso}(z)} \ \times \
\Phi_{obs}[L_B(M_{BH},z),z] \nonumber \\
&& \hskip 1 cm \times \ \frac{d L_B}{d M_{BH}} (M_{BH},z).
\label{rbhobs}
\end{eqnarray}

In figure~\ref{lisa} we plot SMBH formation rates for different masses using
these two previous methods. The solid curves are computed from equation
\ref{rbh2}, i.e. our fiducial model,
and the short-dashed curves are results from equation \ref{rbhobs}. The
different masses considered are labelled in the figure. It can be
seen that there is a significant discrepancy at high redshifts between 
the double-power law predictions and our model's. 
If, as discussed previously (section 4), the efficiency $\epsilon$ declines 
beyond $z \sim 4$, a steeper decline in the SMBH formation
rates at high redshifts would be obtained. This would of course
bring the two estimates in better agreement. 
In other words, this would mean that our model only gives the maximum limiting 
value of SMBH formation rates at these redshifts. 

In any case, it should be noted from the figure that the formation 
rates corresponding to $10^6$ and
$10^7 \ M_\odot$ black holes are predicted to peak at very 
high redshifts ($z \gtrsim 10$) and have
greater values (at least a couple of orders of magnitude larger) than their
more massive counterparts. Therefore, their
detection (or lack of detection) by LISA would yield strong constraints on our
understanding of the formation history of SMBH and 
their plausible minimum masses.

Figure \ref{cumbhrate} shows the cumulative formation rates of SMBH
computed by integrating equation \ref{rbh2} from a minimum black hole mass
$M_{min}$ to a tentative upper limit of $\sim 10^{11}M_\odot $. 
Note that the decline in the cumulative formation
rates of SMBH towards low redshifts is much steeper as compared to the drop in the
cumulative formation rates of haloes (figure \ref{cumformrate}). 

The number density evolution of the quasars with SMBH 
masses above some minimum mass $M_{min}$ can now be written as
\begin{eqnarray}
N_{QSO} ( M_{BH} > M_{min}, z )&=& \int_{M_{min}}^{M_{max}} d M_{BH}
\times \{ \ t_{qso}(z) \nonumber \\  
&& \ \ \  \times \ R_{BH} ( M_{BH},z)\ \}.
\label{nqso} 
\end{eqnarray}
Figure \ref{NQSO} shows the the number density evolution of
quasars with minimum black hole masses as labelled in the figure. 
Also plotted is the observed data for the space density evolution of
quasars with minimum B-band magnitudes $M_B < -24$ (filled circles) and
$M_B < -26$ (open triangles) up to a redshift of 10. 
As can be seen, the figure sketches a consistent picture with the original
estimate of minimum quasar black hole masses of 
$\sim 10^8 M_{\odot}$ by \citet{soltan}. 
\begin{figure}
\hskip 5pt
\epsfxsize 8.0 cm
\epsfbox{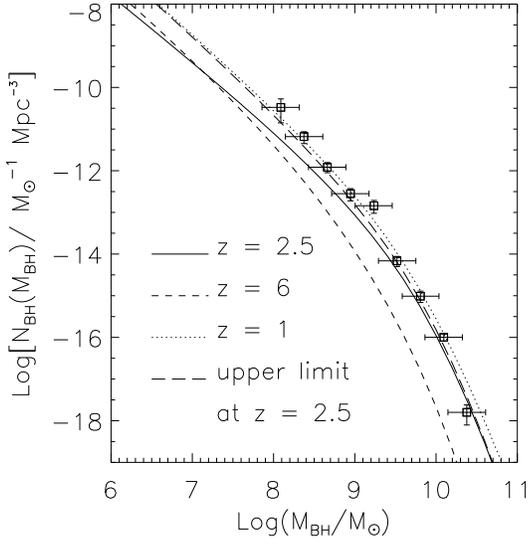}
\caption{ The mass function of SMBHs. Open squares show the local estimates
  given in \citet{saluccietal}. The limiting $z = 2.5$ mass functions from
  our model have been plotted as solid line (using $N_{BH}^A$) and
  long-dashed line (using $N_{BH}^B$). We also show the upper limit 
  mass function at $z=1$ (using $N_{BH}^B$) and at $z=6$, where both limiting
  mass functions ($N_{BH}^A$ and $N_{BH}^B$) are very similar.}
\label{bhmf}
\end{figure}
\begin{figure}
\hskip 5pt
\epsfxsize 8.0 cm
\epsfbox{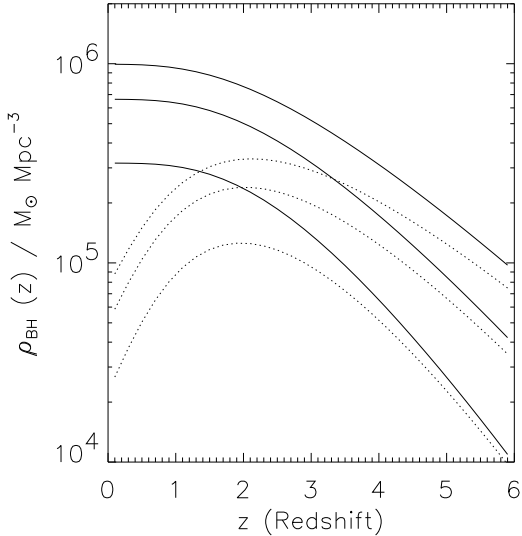}
\caption{Comoving cosmological mass density of SMBHs by using the
 SMBH mass function $N_{BH}^A$ (dotted lines) and $N_{BH}^B$ (solid
 lines). Minimum SMBH masses $M_{min}$ are (top to bottom) $10^8$, $3 \times
 10^8$ and $10^9 \ M_{\odot}$. The two curves (solid and dotted)
 can be considered as (upper and lower) limits on densities for each $M_{min}$. }
\label{rhobh}
\end{figure}
\section{The Mass function and cosmological mass density of SMBHs}

The formation rates of SMBHs can be used to compute their formation epoch
distribution (see section 2.1). These can then be integrated over 
redshift to yield the SMBH mass
function. The formation epoch distribution of SMBHs can be written as 
\begin{eqnarray}
F_{BH}^A ( M_{BH}, z_f \rightarrow z) &=& R_{BH} ( M_{BH},z_f) \
\times \frac{d t}{d z_f}(z_f) \label{fbh1} \\ 
&&\hskip -0.2 cm \times P_{surv}[M_{halo}(M_{BH},z_f), z_f \rightarrow z]
\ . \nonumber  
\end{eqnarray}
However, there are two problems with this equation. The first is 
that the survival probabilities of SMBHs are only crudely correlated to 
the halo survival probabilities: the declining merger rates of SMBH 
in haloes with mass above $M_*$ suggest that the survival 
probability may be greater for SMBHs in these haloes. 
The second problem is more fundamental 
and lies in the fact that equation \ref{mbhz} is not only a function 
of halo mass but also of formation redshift. 
This can be understood in the following way:
masses of SMBH for a given mass halo will be higher
at high redshifts (this holds even if the formation efficiency is
taken as constant such that $\epsilon(z) = \epsilon_0$), so 
that when we compute the destruction probability $P_{dest}$ of 
objects (note that $P_{surv} = 1 - P_{dest}$), we authorise
un-realistic situations where a halo with a higher mass black hole 
goes on to form a lower mass black hole. Hence we underestimate the survival
probability. Equation \ref{fbh1}, therefore, only gives a lower limit
on the number counts of objects of different masses. An upper limit 
can be set by taking $P_{surv} = 1$. This amounts
to computing mass functions by simply integrating the formation 
rates up to the redshift of observation. Such an approach has also been
considered in \citet{haehneltrees}. Thus we write 
 \begin{equation}
F_{BH}^B ( M_{BH}, z_f \rightarrow z) = R_{BH} ( M_{BH},z_f)\times
\frac{d t}{d z_f}(z_f), \ z_f \geq z 
\label{fbh2}.
\end{equation}
We denote the corresponding SMBH mass functions as
$N_{BH}^A(M_{BH},z)$ and $N_{BH}^B(M_{BH},z)$ respectively. These are
computed as
\begin{equation}
N_{BH}^{A,B}(M_{BH},z) = \int_{\infty}^{z} d z_f \ F_{BH}^{A,B} (
M_{BH}, z_f \rightarrow z)\ .
\end{equation}
In figure \ref{bhmf} we plot the SMBH mass functions for
several redshifts and compare them with the estimates of 
local SMBH mass function as given in \citet{saluccietal}. 
At redshifts greater than $z \sim 3$ there appears to be no 
significant difference in the two limiting mass functions 
($N_{BH}^A$ and $N_{BH}^B$). 
We can now write the comoving cosmological SMBH mass density as 
\begin{equation}
\rho_{BH}^{A,B}(M > M_{min},z) = \int_{M_{min}}^\infty \hskip -0.3 cm
d M_{BH} \ \{ M_{BH}
\times N_{BH}^{A,B} (M_{BH},z) \}
\end{equation}

Figure~\ref{rhobh} shows the redshift evolution of
cosmological mass density of the SMBH with minimum masses $ 10^8$, $3
\times 10^8$ and $10^9 \ M_\odot$ (top to bottom) respectively. The
solid lines are computed using $N_{BH}^B$ and dotted lines are
computed using $N_{BH}^A$. It can be seen that mass function
$N_{BH}^A$ gives an unrealistic decline in the mass density of SMBH
which should increase monotonically. Once again, the reason 
is that the survival probability of
the SMBHs are underestimated and thus the number counts are underestimated 
as well\footnote{We point out that in a recent 
paper \citet{bromleyetal} have found a similar non-monotonic evolution 
of SMBH mass density, computed by using the $M-v_c$ relation combined 
with the Press-Schechter mass function.}. The density computed using
the mass function $N_{BH}^B$ thus sets an upper limit to the density
of SMBHs for the three different minimum masses.
\begin{figure*}
\hskip 5pt
\epsfxsize 12.0 cm
\epsfbox{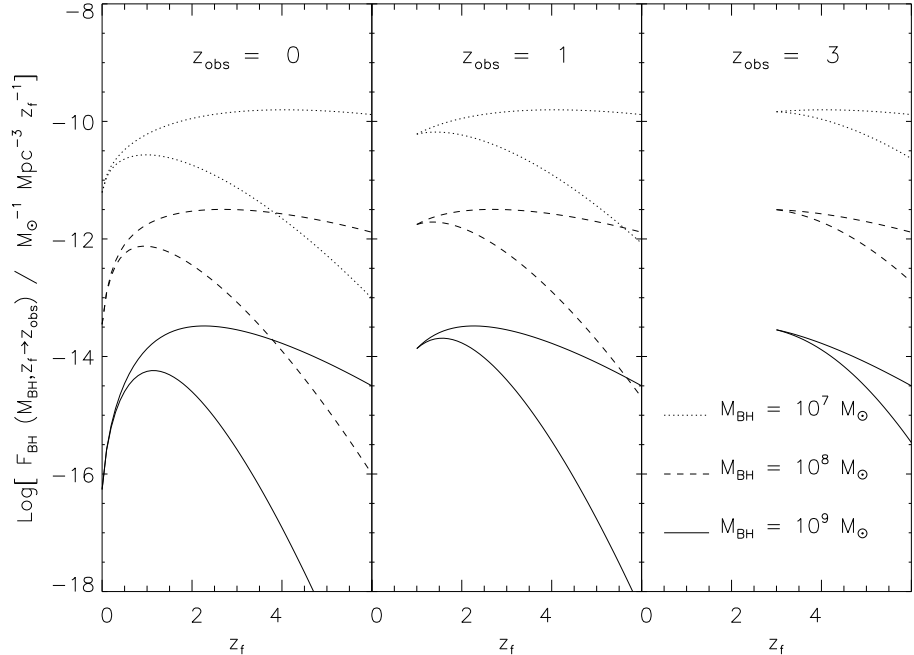}
\caption{The contribution to the present day $z =0$ relic black holes
  from merger activity at higher redshifts. Different line-styles show results of the
  model for different black hole masses whose values are indicated in the figure. 
  Each pair of curve in the same line-style indicates estimates using $F_{BH}^A$ (lower curve) and
  $F_{BH}^B$ (upper curve) as detailed in the text.
 It can be seen that at $z=0$, a significant contribution comes from $z \geq 1$.}
\label{fepochbh}
\end{figure*}

In section 3 we have presented a scenario with a 
monotonically declining efficiency of black hole formation with 
decreasing redshift, with $M_{BH}-v_c$ relation written as
$M_{BH} = \epsilon (z) \ v_c^5$. However, the recent results of
\citet{shieldsetal} point towards a redshift independence of 
$M_{BH} - \sigma$ relation (and therefore, by extension, 
$M_{BH}-v_c$ relation). They find no discernible 
change in the relation up to redshift of $z
\approx 3$. Thus they suggest that black holes are well formed by $z
\approx 3$. In order to reconcile the redshift dependence of our 
model $M_{BH}-v_c$ scaling with observed redshift independence
of this relation, we present an argument based on the formation epoch
distribution of the SMBH. In figure \ref{fepochbh} we plot
the formation epoch distribution of SMBH of different masses at
different redshifts of observation $z_{obs}$. It is evident from the
figure that most relic SMBH between redshifts $z=0$ and $z =1$ 
were formed at high redshifts $z
\gtrsim 1$. As per our model, at redshifts $z \gtrsim 1$, the function
$\epsilon (z)$ does not vary significantly as $z \approx 1$ is a
characteristic redshift of decline for $\epsilon (z)$. Appreciable
decline in $\epsilon (z)$ occurs only at $z \lesssim 1$. Hence, our
adopted redshift varying $M_{BH}-v_c$ relation does not pose a problem
with the observed redshift independence of the relation. 
Furthermore, it can be seen from figure \ref{fepochbh} that using 
$F_{BH}^B$ our model predicts that most relic SMBH between $z=0$ and $z=1$, 
with masses above $10^8\ M_{\odot}$, are formed between $z \sim$ $2$ and $3$.  
Whereas, using $F_{BH}^A$ we predict that most SMBH were formed at $z \sim 1$. 

\section{Estimates of Star Formation Rate and neutral 
gas density in the universe}

In our model, the declining black hole formation efficiency
is attributed to a corresponding decline in the amount of cold
gas mass available in galaxies. To estimate this cold
gas mass, we take the fraction of cold gas
$f_{gas}(z)$ (which is different from the post merger fraction 
of cold gas available for accretion - $f_{cold}$) in galaxies 
as proportional to $\epsilon (z)$, i.e.
$f_{gas} (z) \propto \epsilon (z)$. For simplicity we assume that this
fraction is independent of halo masses. We fix the constant of 
proportionality to match the amount of cold gas observed 
at high redshift in Damped--Ly-$\alpha$ systems. 
The cooling of gas in massive haloes may be ineffective 
\citep{bensonetal} and therefore massive haloes may be deficient in cold gas. 
We, therefore, tentatively adopt an upper limit to the halo masses 
which are rich in cold gas as $M_{max} \sim 10^{14} M_{\odot}$, 
corresponding to cluster scales.

Using these simple prescriptions, we can write the mass of cold gas as 
$M_{gas} (M,z) \ \propto \  \epsilon (z) M$, where $M$ is the halo 
mass. The comoving cosmological mass density of cold gas can then be
written as 

\begin{equation}
\rho_{gas} (z) =  \int_{M_{min}}^{M_{max}}\hskip -0.2 cm d M \ \{M_{gas}(M,z) \times N_{PS}(M,z)\}.
\label{rhogas}
\end{equation}

We take $M_{min}$ as the minimum halo mass corresponding to the
minimum halo circular velocity $v_{m}$ for effectively retaining gas in
the halo potential well \citep{somervilleprimackfaber}. To simplify
our computation we calculate $M_{min}$ corresponding to $v_{min}$ 
using equation \ref{vc} by setting the redshift of formation as the 
redshift of observation itself. This approach has been adopted
previously, for example, in \citet{robinsonsilk}, where they have
discussed the error introduced by this simplification. 
Alternatively one can use the formation epoch
distribution and convert it from masses to circular velocities and
then integrate over all redshifts up to the redshifts of observation
to get the circular velocity functions. 

\begin{figure}
\hbox{\epsfxsize 8.5 cm
\epsfbox{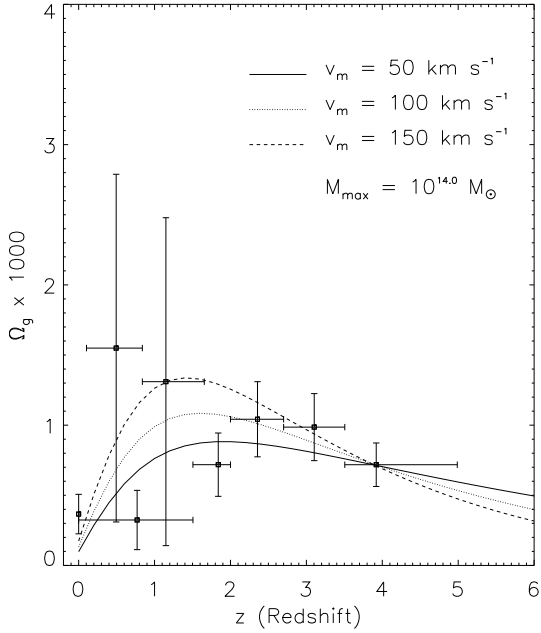}}
\caption{The evolution of comoving cosmological mass density of cold
  gas in galaxies as discussed in section 7. The data points are as
  given in \citet{perouxetal}, and the different line-styles correspond to 
estimates truncated at different low circular velocities indicated on the
  figure (see text for details). }
\label{gas}
\end{figure}

Dividing $\rho_{gas}$ by the critical density $\rho_c$ of 
matter in the universe we obtain $\Omega_{gas}$. This is plotted in 
figure \ref{gas} against the observations of Damped--Ly-$\alpha$
absorbers \citep{perouxetal}. As previously mentioned, we have normalised 
$\Omega_{gas}$ to the highest redshift data point in figure \ref{gas}
which yields a constant of proportionality of the order of $10^{-1}$ 
for the $f_{gas} (z) - \epsilon (z)$ relation, 
with the exact value depending on the lower circular velocity 
cut-off used in the computation of $\Omega_{gas}$.

Further, the star formation rates can be estimated by adopting an SFR recipe as
(e.g. Sommerville \& Primack 1999)
\begin{equation}
\dot{M}_* = \epsilon_* \frac{M_{gas}}{t_*},
\end{equation}
where $\epsilon_*$ is the efficiency of star formation and $M_{gas}$
is the cold gas mass available for star formation as described above. We can thus write the SFR
density in the universe as
\begin{equation}
\rho_{SFR} (z) = \int_{M_{min}}^{M_{max}} d M \ \dot{M}_*  \times N_{PS}(M,z) .
\label{SFRint}
\end{equation}
\begin{figure}
\hbox{\epsfxsize 8.7 cm
\epsfbox{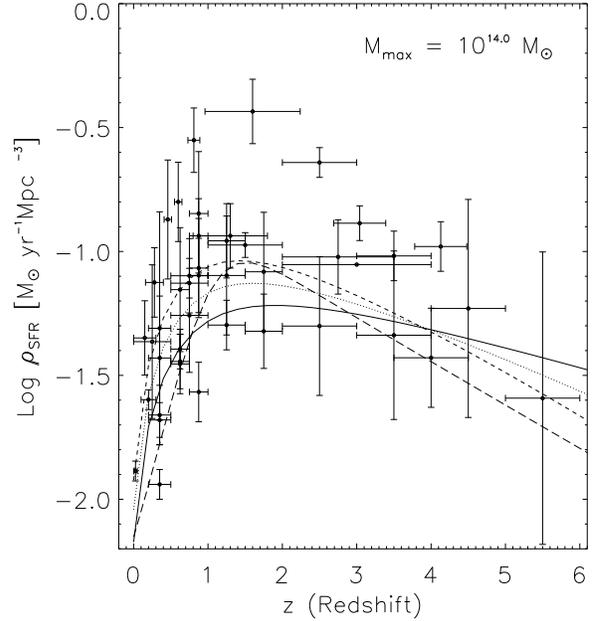}}
\caption{The evolution if SFR density in the universe. The different curves 
  are results of the model with different minimum 
  circular velocities using the same line-style code as in figure \ref{gas}. 
  The long-dashed curve is from the empirical SFR
  density evolution given in \citet{madaupozzetti}. The data points with error
  bars are as compiled in \citet{ascasibaretal}.}
\label{sfrd}
\end{figure}

Retaining the same normalisation as for the cold gas density, we 
choose $t_* = 10^8$ years and an efficiency of 5 percent i.e. $\epsilon_*
= 0.05$. $M_{min}$ and $M_{max}$ are as discussed for the case of
cold gas density. Figure \ref{sfrd} shows the evolution of the SFR
density in the universe.

It can be seen from the figure that our crude estimates for the cold
gas density and the SFR density are broadly consistent with the data.
This supports our assumption that there exists a correlation between the
declining cold gas mass with redshift and the declining efficiency 
of black hole formation with redshift.
\begin{figure*}
\hskip 5pt
\epsfxsize 14.0 cm
\epsfbox{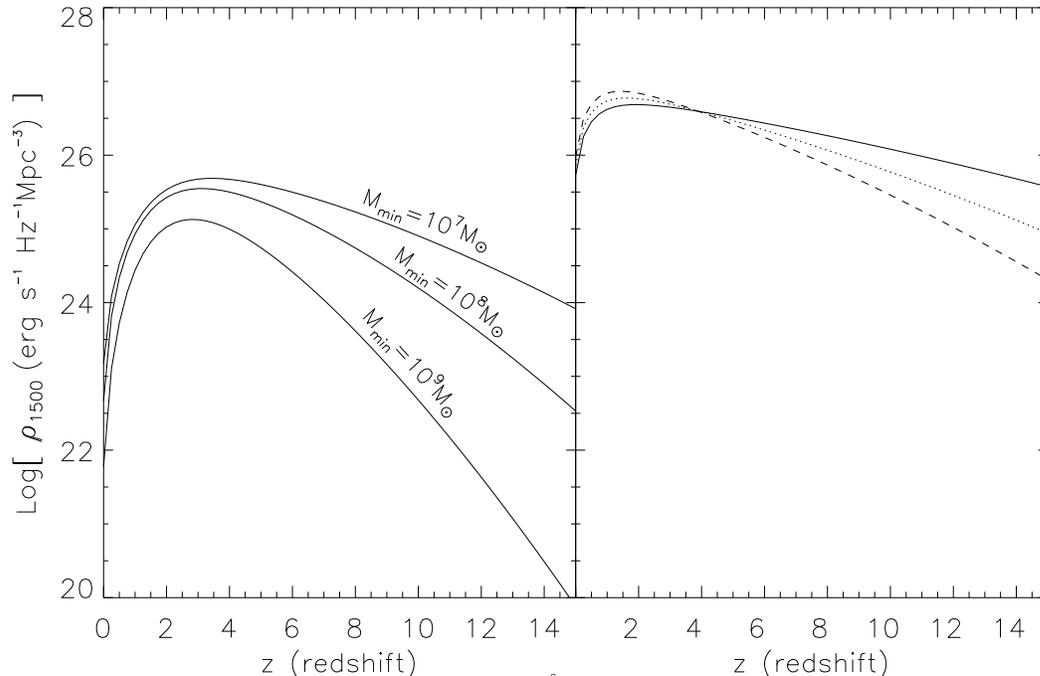}
\caption{\it{Left panel:} \rm The evolution of luminosity
  density (in 1500 $\AA$) for quasars with minimum black 
  hole masses $M_{min}\ = \ 10^7, \ 10^8$ and $10^9 \ M_{\odot}$ 
  respectively. \it{Right panel:} \rm The evolution of luminosity
  density (in 1500 $\AA$) estimated from the star formation rates 
  in the universe (see text for details). The line-styles are same as
  in figure \ref{gas}. }
\label{ionback}
\end{figure*}

\section{Simple estimates of the quasar and the star formation
 luminosity density in the universe}

Extending the analysis of SFR density and cold gas density in the
universe, we now compute the the relative contribution to the ionising 
background arising from the star formation rates and 
quasars. In the previous section we computed star formation rates
by assuming a direct proportionality between the fraction of cold gas in
galaxies and the efficiency of black hole formation. In a more realistic
scenario, the cold gas content in galaxies builds up gradually, and is
controlled primarily by two competing processes. The first process
is that of gas condensation, whereby the gas loses energy over a cooling time
scale and falls into the central baryonic core over a dynamical infall
time-scale to form stars. The second process involves the expulsion of a
fraction of this gas from
the galaxy as a result of feedback from supernovae and AGNs. It
seems reasonable, therefore, to suppose that the gas
content in galaxies evolves dynamically, peaking at some epoch and
declining afterwards. 

However, in our model the black hole formation efficiency does not
peak at high redshifts but reaches a maximum saturating
value. Nevertheless, in view of the consistent results for the
evolution of $\Omega_{gas}$ and $\rho_{SFR}(z)$, we bypass the exact 
evolution history of $\epsilon (z)$ and $f_{gas}(z)$ by assuming their 
parallel evolution. This way we compute the maximal
evolution of the luminosity densities for star formation rates and the
quasars by extrapolating the model $\epsilon (z)$ to high redshifts. 
It is possible that at high redshifts, the evolution of star
formation rates as well as the black hole formation rates are 
driven predominantly by a rapid gravitational growth of structure
\citep{dimatteoetal}. Therefore, at these redshifts the 
exact evolution of $\epsilon (z)$ and $f_{gas}(z)$ is perhaps not so
important.

Now the ultraviolet luminosity arising from
star formation can be written as \citep{madauetal}
\begin{equation}
L_{UV} = \rm C \times (\ SFR/ \ \it M_{\odot} \rm yr^{-1})\  erg \
s^{-1} Hz^{-1}.
\end{equation}
At $1500 \ \AA$, just longward of the hydrogen Lyman limit (e.g. HNR98), 
the value of $\rm C$ for a Salpeter IMF is given as 
$\rm C \approx 8 \times 10^{27}$. Thus the ultraviolet star formation 
luminosity density is written as
\begin{equation}
\rho_{1500} = \rm C \it \times \rho_{SFR}.
\end{equation}

For quasars with a minimum black hole mass as $M_{min}$, 
the luminosity density in $B$-band can be written as
\begin{eqnarray}
\rho_{4400}(z) &=& \int_{M_{min}}^{\infty} d M_{BH} \times 
[ \ \frac{L_B(M_{BH})}{\ \nu_{B}}  \\ \nonumber
&&\hskip 1 cm \times  \ t_{qso}(z) \times R_{BH}(M_{BH},z) \ ].
\end{eqnarray}
Here, $t_{qso}(z) \ \times \ R_{BH}(M_{BH},z)$ is the number of
active quasars with black hole mass $M_{BH}$ existing at 
redshift $z$. To convert the $B$-band luminosity into luminosity at 
$1500 \AA$, we adopt a power law spectral energy distribution 
scaling as $L_{\nu} \propto \nu^{-\alpha}$ \citep{pei}, 
where $\alpha = 0.5$ and $L_{\nu}$ for a given band is simply 
$L_{band}/\nu_{band}$. In figure \ref{ionback}, we plot the evolution
of luminosity densities due to star formation rates and quasars. These
luminosity densities have also been estimated previously, for example,
in \citet{boyleterlevich}. For minimum black hole masses in the range
$10^7$ to $10^9$, we find that the drop in the quasar luminosity density
towards low redshifts is steeper than computed in \citet{boyleterlevich}.  
For minimum quasar black hole
masses of $10^8 M_\odot$ the curves in fig \ref{ionback} suggest that
at high redshifts, the ionising background is predominantly of stellar
origin. It comes out from our calculations that quasar dominated 
ionising photon
background is plausible if the minimum quasar black hole masses are
$\lesssim 10^7 M_\odot$ and the minimum circular velocities for haloes
to effectively retain gas are fairly high ($\sim 150 \ \rm km \
s^{-1}$). However, in light of the simplicity of the model used and the
uncertainties involved in each of the estimates, these results should only be 
considered as indicative.   

\section{discussion and conclusions}

We have presented an empirically motivated model for accretion
dominated self-regulated growth of super massive black holes in galaxies and
investigated their implications for the evolution of the quasar
population in the universe. The model looks into the core aspects of
quasar population such as the space density evolution of quasars, the
evolution of the quasar characteristic luminosity, plausible minimum
masses of quasars, the mass function of the super massive
black holes and the formation epoch distribution of these black holes. 

The model suggests that the characteristic luminosity in
the quasar luminosity functions arises primarily as a consequence of a
characteristic mass-scale above which there is a systematic deviation
of the black hole merging rates from the halo merger rates. Up to this 
mass-scale, the black hole merging appears to be closely
following the merging of corresponding dark matter haloes. This
anomalous behaviour is probably a consequence of rarer galaxy merging
in massive haloes. Galaxies in massive haloes tend to
retain their properties rather than quickly sink to the 
centre through dynamical friction. The extent of this effect
 might depend on the
local environment of the merging galaxies. It is possible, for
example, that galactic merging in massive haloes is efficient for 
high sigma peaks only. We have mimicked such a behaviour by 
introducing a simple phenomenological 
decline in the black hole formation rate through a function $F_\mu (M)$ 
with a power law drop at the halo high mass end.
When combined with a declining efficiency of
black hole formation with redshift which we attribute to a decrease in the
cold gas mass available, the model can reproduce the quasar
luminosity function for a wide range of redshifts. Furthermore,
we have shown that this declining black hole formation 
efficiency gives a better comparison of
the model break luminosity evolution with the evolution of
characteristic luminosity used in the empirical double power-law 
fit to the data.

On the lines of WL03, we have modelled time-scales for quasars 
proportional to the local dynamical time of the newly forming baryonic 
core, which in turn is proportional to the halo dynamical time. 
We found that the local dynamical time gives a 
better comparison with low redshift observations than the Salpeter time-scale. 
Furthermore, the adopted redshift varying time-scale allows a natural translation
of the black hole formation rate density into number density of quasars, consistent 
with the the observed data. 

Calculation of the number density evolution of quasars in our model also led us
to infer plausible minimum masses of optical quasars on the order of
$\sim 10^8 M_{\odot}$, consistent with the
estimates of \citet{soltan} who computed minimum masses assuming a
radiative efficiency of optical quasars  $\epsilon_{rad} = 0.1$. 
Additionally, our model suggests a less steep decline of quasar number density at
high redshifts than is generally assumed. 

We also calculated the evolution of the comoving density
of black holes with masses above $10^8 M_\odot$ and $\rm 3 \times
10^8$ and $10^9 \ M_{\odot}$. The possible reasons for anomalous 
non-monotonic evolution of $\rho_{BH}$ inferred from the black hole
mass function $N_{BH}^A$ have been discussed. 
We find that for minimum black hole masses of
$10^9 M_{\odot}$ a density of at least $\rho_{BH} \sim 10^5 M_\odot \rm
Mpc^{-3}$ has assembled by redshift of 2. The absolute upper limit
for minimum SMBH masses of $10^8 M_{\odot}$ is $\rho_{BH} \sim 10^6
M_\odot \rm Mpc^{-3}$. These estimates are consistent with the 
estimates of local black hole density (Chokshi \& Turner 1992;
Salucci et al. 1999; Yu \& Tremaine 2002). Our model suggests that 
a significant fraction of SMBH mass density was already in place by 
$z \sim \rm 2 \ to \ 3$. 
Interestingly, these estimates are also consistent with
the recent results of \citet{dimatteoetal}. Moreover, the model SMBH mass
functions at $z \lesssim 2.5$ are consistent with the local mass function
of SMBH derived in \citet{saluccietal}. 

Finally, the predicted ionising background at high redshift appears to be
predominantly of stellar origin. We find that a quasar dominated
ionising background would need a lower limit of minimum SMBH
masses along with a higher value of minimum circular velocities of
haloes to effectively retain gas in them. Obscured low mass AGNs could
thus raise the prospects of a quasar dominated ionising
background. However, it is also possible that low mass AGNs have relatively
low radiative efficiencies (e.g. Yu \& Tremaine 2002) and thus may not
be able to contribute effectively to the ionising background.

\section*{acknowledgements}
A. M. wishes to thank Sugata Kaviraj, Ranty Islam, Claus Beisbart and Steve Rawlings
for their help during the course of this work. Also, many thanks to Jeremy
Blaizot for discussions, suggestions and his optimism
about the present work. A. M. gratefully acknowledges the support from
the Eddie Dinshaw foundation at Balliol College, Oxford. The
research of J. E. G. D. at Oxford was supported by a major grant of the 
Leverhulme foundation.

\appendix

\section{A note on the halo major merger rates}

The halo major merger rates derived in section (2.1) are based on the
approach of Kitayama \& Suto (1996) whereby the major mergers are
characterised as merger events in which an object at least doubles its
mass. As pointed out in section 2, in this definition of major merger events, 
there is always a chance that the main progenitor has more than half the
mass of the final halo, even though it is not supposed to. We discuss here how one
can systematically account for this possibility and compute
corrected major merger rates where no progenitor has more than half
the mass.

Consider a merger event in which an object of mass $M$ is formed from a
progenitor of mass $M_1$ ($M_1 < M/2$). The mass accreted by this
progenitor is $M - M_1 = \Delta M$ ($\Delta M > M/2$). This mass
$\Delta M$ can come either as a combination of formed smaller
mass objects $m_i$ such that $\sum_i m_i = \Delta M$ or 
as a single formed object of mass $\Delta M$. In order to
distribute the mass $\Delta M$ (variance denoted as $S_{\Delta}$) 
into smaller mass objects of mass $m$ (variance denoted as $S_m$)
we use the Press-Schechter mass function (at time $t$ corresponding to
$\omega \approx \omega_1 \approx \omega_2$ ). In the excursion set
picture of \citet{bondetal}, the random walks of trajectories $\delta
(S_m)$ for the sub-clumps $m$ now begin at 
$S_m \ = \ S_\Delta$ instead of $S_m \ =  \ 0$. This simply leads to 
an offset in the first passage distribution function, 
which can be written as
\begin{equation}
f(S_m|S_\Delta) d S_m = \frac{1}{\sqrt{2 \pi}} \frac{\omega}{(S_m -
  S_\Delta)^{3/2}}  \rm exp \it \left[ - \frac{\omega^{\rm 2}}{ \rm 2
  \it (S_m -
  S_\Delta)} \right] \ dS_m.
\label{progenitor}
\end{equation}
The probability that none of the masses $m_i$ is greater than
$M/2$ can be inferred from the `fraction' of mass of $\Delta M$
existing in objects below mass $M/2$. Thus 
\begin{equation}
P_m[m < (M/2)\ |\ \Delta M] = \rm erf \it \left[\frac{\omega}{\sqrt{\rm 2 \it (S_h - S_\Delta)}}\right],
\end{equation}
where $S_h = S(M/2)$. Note that when $S_\Delta \ \rightarrow \ S_h$ this probability is one,
that is all the masses $m_i$ are less than $M/2$, as expected. Hence the corrected major
merger rate is given as
\begin{eqnarray}
R_{MM}(M,t) &=& N_{PS}(M,t)\int_{S_h}^\infty d S_1 \ \frac{d P_1}{d t} \
P_m \label{AppendRM} \\ \nonumber
&=& N_{PS}(M,t) \left[-\frac{d \omega}{dt}\right] I(M,t)
\end{eqnarray}
where $I(M,t)$ is given as
\begin{equation}
I(M,t) =  \int_{S_h}^\infty dS_1 \ \frac{1}{\sqrt{2 \pi}} \frac{1}{[S_1 -S(M)]^{3/2}}
\ \rm erf \it \left[\frac{\omega(t)}{\sqrt{\rm 2 \it
      (S_h - S_\Delta)}}\right].  
\end{equation}
Unlike the expression for merger rates given in KS96, 
we find that $I(M,t)$ here is a function of
time. For any given mass $M$, therefore, KS96's expression becomes more
accurate as redshift increases (or for large $\omega$) as $P_m
\rightarrow 1$. Unfortunately the full expression for $I(M,t)$ is not easy to
evaluate. The difference becomes more pronounced at low redshifts, 
even though for massive haloes ($ > 10^{12}M_\odot$ at $z=0$) the discrepancy is
too weak to affect the results presented in this paper. Hence for the purpose
of deriving quasar luminosity functions and other properties we
have restricted ourselves to the `computationally easy' expression of KS96.

An additional advantage of equation~\ref{progenitor} is that it 
could be used in the merger trees to sub-divide the remaining mass 
after picking the main progenitor at an earlier time in a short
time-step. If this time step is very small, then in most cases the 
remainder mass is small and can be treated as accreted. In other cases, 
 we think it more satisfactory to distribute the remaining mass 
(other than the main progenitor) according to equation \ref{progenitor}
rather than use the various other methods that have been proposed in the literature.


\begin{thebibliography}{}
\bibitem[\protect\citeauthoryear{Ascasibar et al.}{2002}]{ascasibaretal} 
Ascasibar Y., Yepes G., Gottlober S., Muller V. \AAA{387}{396}{2002} 
\bibitem[\protect\citeauthoryear{Efstathiou, Bond \& White}{1992}]{ebw} 
Efstathiou G., Bond J. R., White S. D. M.,\MNRAS{258}{1p}{1992} 
\bibitem[\protect\citeauthoryear{Benson et al.}{2003}]{bensonetal} 
Benson A. J., Bower R. G., Frenk C. S., Lacey C. G., Baugh C. M., Cole
S., \astroph{0302450} 
\bibitem[\protect\citeauthoryear{Binney \& Tremaine}{2002}]{binney} 
Binney J., Tremaine S., Galactic Dynamics, Princeton University Press, 1987 
\bibitem[\protect\citeauthoryear{Bond et al.}{1991}]{bondetal} 
Bond J. R., Cole S., Efstathiou G., Kaiser N. \ApJ{379}{440}{1991} 
\bibitem[\protect\citeauthoryear{Bower}{1991}]{bower} 
Bower R. J. \MNRAS{248}{332}{1991} 
\bibitem[\protect\citeauthoryear{Boyle \& Terlevich}{1998}]{boyleterlevich} 
Boyle B. J., Terlevich R. J. \MNRAS{293}{49}{1998} 
\bibitem[\protect\citeauthoryear{Boyle et al.}{2000}]{boyleetal} 
Boyle B. J., T. Shanks, S. M. Croom, R. J. Smith, L. Miller,
N. Loaring, and C. Heymans, \MNRAS{317}{1014}{2000}
\bibitem[\protect\citeauthoryear{Bromley, Somerville \& Fabian}{2003}]{bromleyetal}
Bromley J. M., Somerville R. S., Fabian A. C., \astroph{0311008}
\bibitem[\protect\citeauthoryear{Bryan \& Norman}{1998}]{bryannorman}
Bryan G., Norman M. L., \ApJ{495}{80B}{1998}
\bibitem[\protect\citeauthoryear{Burkert \& Silk}{2001}]{burkertsilk} 
Burkert A.., Silk J., \ApJ{151}{154}{2001}  
\bibitem[\protect\citeauthoryear{Cavaliere \& Vittorini}{2000}]{cavalierevittorini} 
Cavaliere A., Vittorini V., \ApJ{543}{599}{2000} 
\bibitem[\protect\citeauthoryear{Chokshi \& Turner}{1992}]{chokshiturner} 
Chokshi A., Turner E. L., \MNRAS{259}{421}{1992} 
\bibitem[\protect\citeauthoryear{Di Matteo et al.}{2003}]{dimatteoetal} 
Di Matteo T., Croft R. A. C., Springel V., Hernquist L., \ApJ{593}{56}{2003}
\bibitem[\protect\citeauthoryear{Fan et al.}{2001}]{fanetal} 
Fan et al., \AJ{121}{54}{2001}
\bibitem[\protect\citeauthoryear{Ferrarese \& Merritt}{2000}]{ferraresemerritt} 
Ferrarese L., Merritt D., \ApJ{539}{9}{2000}
\bibitem[\protect\citeauthoryear{Ferrarese}{2002}]{ferrarese} 
Ferrarese L., \ApJ{578}{90}{2002}
\bibitem[\protect\citeauthoryear{Gebhardt et al.}{2000}]{gebhardtetal} 
Gebhardt K. et al., \ApJ{539}{13}{2000}
\bibitem[\protect\citeauthoryear{Haehnelt \& Rees}{1993}]{haehneltrees} 
Haehnelt, M., Rees, M., \MNRAS{263}{168}{1993}
\bibitem[\protect\citeauthoryear{Haehnelt, Natarajan \& Rees}{1998}]{haehneltnatarajanrees} 
Haehnelt, M., Natarajan, P., Rees, M., \MNRAS{300}{817}{1998} (HNR)
\bibitem[\protect\citeauthoryear{Haiman, Ciotti \& Ostriker}{2003}]{haimanciottiostriker} 
Haiman, Z., Ciotti L., Ostriker J. P., \astroph{0304129}
\bibitem[\protect\citeauthoryear{Hatziminaoglou et al.}{2003}]{hatiziminaoglouetal}  
Hatziminaouglou, E., Mathez, G., Solanes, J. M., Manrique,
  A., Sole, E. S., \MNRAS{343}{692}{2003} 
\bibitem[\protect\citeauthoryear{Hosokawa}{2002}]{hosokawa} 
Hosokawa T., \ApJ{576}{75}{2002}
\bibitem[\protect\citeauthoryear{Kauffmann \& Haehnelt}{2000}]{kauffmannhaehnelt} 
Kauffmann G., Haehnelt M., \MNRAS{311}{576}{2000} 
\bibitem[\protect\citeauthoryear{Kennefick, Djorgovski \& Carvalho}{1995}]{kennefick} 
Kennefick, J. D., Djorgovski, S. G., de Carvalho, R. R., \AJ{110}{2553}{1995}
\bibitem[\protect\citeauthoryear{Kitayama \& Suto}{1996}]{kitayamasuto} 
Kitayama T., Suto Y., \MNRAS{230}{638}{1996}
\bibitem[\protect\citeauthoryear{Lacey \& Cole}{1993}]{laceycole} 
Lacey C., Cole S., \MNRAS{262}{627}{1993}
\bibitem[\protect\citeauthoryear{Lacey \& Cole}{1994}]{laceycole2} 
Lacey C., Cole S., \MNRAS{271}{676}{1994}
\bibitem[\protect\citeauthoryear{Madau, Pozzetti \& Dickinson}{1998}]{madauetal} 
Madau P., Pozzetti L., Dickinson M \ApJ{498}{106}{1998}
\bibitem[\protect\citeauthoryear{Madau \& Pozzetti}{2000}]{madaupozzetti} 
Madau P., Pozzetti L., \MNRAS{312}{9}{2000}
\bibitem[\protect\citeauthoryear{Navarro, Frenk \& White}{1997}]{nfw} 
Navarro, J. F., Frenk, C. S., White, S. D. M., \ApJ{490}{493}{1997} 
\bibitem[\protect\citeauthoryear{Pei}{1995}]{pei} 
Pei Y. C., \ApJ{438}{623}{1995}
\bibitem[\protect\citeauthoryear{Peroux et al.}{2001}]{perouxetal} 
Peroux C., McMahon R. G., Storrie-Lombardi L. J., Irwin M. J. \astroph{0107045}
\bibitem[\protect\citeauthoryear{Robinson \& Silk}{2000}]{robinsonsilk} 
Robinson J., Silk J., \ApJ{539}{89}{2000}
\bibitem[\protect\citeauthoryear{Salucci et al.}{1999}]{saluccietal} 
Salucci P., Szuszkiewicz E., Monaco P., Danese L. \PASJ{637}{644}{1999}
\bibitem[\protect\citeauthoryear{Sasaki}{1994}]{sasaki} 
Sasaki S., \PASJ{46}{427}{1994}
\bibitem[\protect\citeauthoryear{Schmidt, Schneider \& Gunn}{1995}]{schmidtschneidergunn} 
Schmidt M., Schneider D. P., Gunn J. E., \AJ{110}{68}{1995}
\bibitem[\protect\citeauthoryear{Shaver et. al}{1996}]{shaveretal} 
Shaver P. A, Wall J. V., Kellermann K. I., Jackson C. A., Hawkins
M. R. S., \Nature{384}{439}{1996} 
\bibitem[\protect\citeauthoryear{Shields et al.}{2003}]{shieldsetal} 
Shields G. A., Gebhardt K., Salviander S., Wills, B. J., Xie B., Brotherton M. S., Yuan J., Dietrich M., \ApJ{583}{124}{2003}
\bibitem[\protect\citeauthoryear{Silk \& Bouwens}{2001}]{silkbouwens} 
Silk J., Bouwens, R., \NewARev{45}{337}{2001}
\bibitem[\protect\citeauthoryear{Silk \& Rees}{1998}]{silkrees} 
Silk J., Rees M. J., \AAA{331}{L1}{1998}
\bibitem[\protect\citeauthoryear{Soltan}{1982}]{soltan} 
Soltan A., \MNRAS{}{115S}{1982}
\bibitem[\protect\citeauthoryear{Somerville \& Primack}{1999}]{somervilleprimack} 
Somerville R. S., Primack J. R., \MNRAS{310}{1087}{1999}
\bibitem[\protect\citeauthoryear{Somerville, Primack \& Faber}{2001}]{somervilleprimackfaber} 
Somerville R. S., Primack J. R., Faber S. M., \MNRAS{320}{504S}{2001}
\bibitem[\protect\citeauthoryear{Warren, Hewett \& Osmer}{1994}]{who} 
Warren S. J., Hewett P. C., Osmer P. S., \ApJ{421}{412}{1994} 
\bibitem[\protect\citeauthoryear{Wyithe \& Loeb}{2003}]{wyitheloeb} 
Wyithe J. S. B., Loeb A., \ApJ{595}{614}{2003}
\bibitem[\protect\citeauthoryear{Yu \& Tremaine}{2002}]{yutremaine} 
Yu Q., Tremaine S., \MNRAS{335}{965}{2002}
\end{thebibliography}
\end{document}